\documentclass[final,5p,times,a4paper,number]{elsarticle}

\usepackage{amssymb}
\usepackage{amsmath}
\journal{Your Journal Name}

\usepackage[T1]{fontenc}
\usepackage[utf8]{inputenc}
\usepackage{textcomp}
\usepackage{amsmath,amssymb}
\usepackage{booktabs}
\usepackage{adjustbox}
\usepackage{placeins}
\usepackage{graphicx}
\biboptions{numbers,sort&compress}
\newcommand{\imgH}{0.55\textheight}
\newcommand{\img}[2][\linewidth]{%
  \IfFileExists{#2}%
    {\includegraphics[width=#1,height=\imgH,keepaspectratio]{#2}}%
    {\PackageWarning{pandoc2journal}{Image not found: #2}%
     \fbox{\parbox[c][5cm][c]{\dimexpr#1-2\fboxsep-2\fboxrule\relax}{\centering\sffamily\small Missing image:\\ \detokenize{#2}}}}}
\usepackage[expansion=false]{microtype}   %% protrusion only: safe with every font
\usepackage{dblfloatfix}

\ifdefined\hypersetup\else
  \usepackage{xcolor}
  \usepackage[colorlinks=true,linkcolor=blue!50!black,citecolor=blue!50!black,urlcolor=blue!50!black,breaklinks=true]{hyperref}
\fi
\providecommand{\texorpdfstring}[2]{#1}

\begin{document}

\begin{frontmatter}



\title{Orientation-Tunable Interfacial Thermal Conductance in Lateral MoSe\textsubscript{2}/MoS\textsubscript{2} Heterostructures: A Deep Potential Molecular Dynamics Study}


\author[a]{Shirin Sabokdast\fnref{equal}}
\author[b]{Pedram Mirchi\fnref{equal}}
\author[a]{Ali Rajabpour\corref{cor1}}
\ead{rajabpour@eng.ikiu.ac.ir}
\fntext[equal]{Shirin Sabokdast and Pedram Mirchi contributed equally to this work.}
\cortext[cor1]{Corresponding author}
\affiliation[a]{organization={Mechanical Engineering Department, Imam Khomeini International University},
            city={Qazvin},
            country={Iran}}
\affiliation[b]{organization={Department of Mechanical Engineering, K. N. Toosi University of Technology},
            city={Tehran},
            country={Iran}}

%% Abstract
\begin{abstract}
Managing heat flow across interfaces in lateral 2D transition-metal dichalcogenide (TMD) heterostructures is a key challenge for next-generation nanoelectronic and optoelectronic devices. To tackle this, we trained a deep learning interatomic potential (DLP) using DeePMD-kit on density functional theory (DFT) data, then paired it with non-equilibrium molecular dynamics (NEMD) simulations to study interfacial thermal conductance (ITC) in lateral MoSe\textsubscript{2}/MoS\textsubscript{2} heterostructures, considering both armchair and zigzag interface orientations. The potential we trained closely matches DFT energies, forces, and phonon dispersion relations, which confirms its reliability for thermal transport simulations. Examining cell lengths from 10 to 40 nm and extrapolating to infinite length, the armchair interface shows a higher ITC (473 MW/m\textsuperscript{2}·K) than the zigzag interface (387 MW/m\textsuperscript{2}·K), and the armchair interface remains higher at every length considered. Both increase further as temperature rises. At 300 K, the vibrational density of states shows greater phonon spectral overlap at the armchair interface than at the zigzag interface. Together, these results offer a solid machine-learning framework for studying heat transport in lateral TMD heterostructures, and identify interface orientation as an effective means of controlling heat dissipation at these junctions.
\end{abstract}



%% Keywords

\end{frontmatter}

\FloatBarrier
\section{Introduction}

TMDs represent a relatively young class of 2D layered materials that
have already attracted considerable attention --- mostly because their
electrical and magnetic behavior is nothing like zero-band-gap graphene
or insulating hexagonal boron nitride~\cite{r1,r2,r3}. Single- and few-layer
forms of MoS\textsubscript{2}~\cite{r2}, MoSe\textsubscript{2}~\cite{r4}, WS\textsubscript{2} and WSe\textsubscript{2}~\cite{r5}, along with heterostructures built from these compounds~\cite{r6,r7}, have already been made in the lab and applied in catalysis, energy
storage, chemical sensing, field-effect transistors, and logic circuits~\cite{r8,r9,r10}. However, device miniaturization concentrates heat
dissipation in ever smaller volumes, making thermal transport a critical
design parameter. Understanding thermal transport in 2D TMDs has
therefore become a priority as these materials move from the laboratory
into devices~\cite{r11,r12}. 

Within the TMD family, the molybdenum-based pair MoX\textsubscript{2} (X = S, Se) is
especially interesting. Both compounds share the same graphite-like
X--Mo--X layered structure, and the structural and lattice-parameter
compatibility of MoS\textsubscript{2} and MoSe\textsubscript{2} makes them a suitable pair for
heterostructure formation~\cite{r13,r14,r15}. One study examined exactly
this, combining first-principles calculations with Boltzmann transport
theory to determine the thermal transport behavior of a MoS\textsubscript{2}/\allowbreak{}MoSe\textsubscript{2}
bilayer heterostructure. At room temperature, a thermal conductivity of
25.39 W m\textsuperscript{$-$1} K\textsuperscript{$-$1} was obtained, lying between the values of monolayer MoS\textsubscript{2}
and monolayer MoSe\textsubscript{2}~\cite{r16}. Broad studies based on first principles on
the semiconducting MX\textsubscript{2} family (M = Cr, Mo, W; X = S, Se, Te) show that
the band gaps range from the visible light spectrum into the
near-infrared, which is well suited for nanoelectronics and
optoelectronics~\cite{r17,r18}. Modeling how atoms interact within
these structures and heterostructures is foundational to understanding
their physical properties, and two established approaches exist. Since
the interatomic forces in ab initio molecular dynamics (AIMD)---which
are calculated on the fly using density functional theory (DFT)---are
determined from first principles, it is possible to achieve
first-principles accuracy; nevertheless, the high computational cost
usually limits the simulations to a few hundred atoms and tens of
picoseconds. This is far below the system sizes (tens of thousands of
atoms or more) and the timescales (nanoseconds) needed for steady-state
NEMD calculations of the interfacial thermal conductance in TMD
heterostructures~\cite{r19,r20}.

Empirical-potential MD (e.g., Stillinger--Weber, Tersoff, REBO)
represents a common alternative approach, since it enables access to
much larger scales, namely hundreds of nanometers and nanoseconds of
simulated time~\cite{r21,r22,r23,r24}. The tradeoff is accuracy. Empirical
potentials tend to miss the finer points of atomic interactions, and
they exhibit notable limitations in predicting energy derivatives such
as forces and virials, which are critical at the nanoscale. They also do
not transfer well: each one is usually fit to a narrow set of
applications, so employing a single potential to capture both ITC and
mechanical properties simultaneously is often unreliable. A method that
accurately reproduces atomic trajectories while retaining computational
efficiency is therefore needed~\cite{r25,r26}. With the growing adoption
of machine learning in materials science, machine-learned approaches
have provided a viable route to overcoming this limitation~\cite{r27}. Out
of that effort came a range of machine-learned interatomic potentials
(MLIPs) --- GAP~\cite{r28,r29}, NEP~\cite{r30,r31}, and neural network
potentials (NNPs)~\cite{r32}, among others. NNPs, in particular, merit
specific attention: with the right architecture, a neural network can
approximate almost any complex, multidimensional function to arbitrary
precision, which makes it a natural fit for mapping atomic positions to
energy. Some studies have even pushed NNPs to quantum-mechanical levels
of accuracy, for both small and extended systems, while still scaling
sensibly with system size~\cite{r33,r34}. The loss function includes
terms related to energy and force, so that the model is trained not
only on the energy itself but also on the derivatives of the potential
energy surface, thereby greatly increasing the amount of information
obtained from each DFT configuration. NNPs have already proven
themselves across a wide range of material systems --- GaN/\allowbreak{}BAs~\cite{r35},
GaN/\allowbreak{}SiC~\cite{r36}, GaN/\allowbreak{}AlN/\allowbreak{}SiC~\cite{r36}, N--Ga--Al~\cite{r26}, ZnO/\allowbreak{}GaN~\cite{r37}, Si/\allowbreak{}Ge~\cite{r38}, MoS\textsubscript{2}/\allowbreak{}WS\textsubscript{2}~\cite{r39}, and single-layer penta-NiN\textsubscript{2}~\cite{r40}, among others --- which
substantiates their status as one of the most versatile and promising
MLIP approaches currently available.

While NNPs have been successfully applied to several TMD-based
interfaces~\cite{r39,r41}, the specific role of interface crystallographic
orientation in governing the ITC of the MoS\textsubscript{2}/\allowbreak{}MoSe\textsubscript{2} system, at
first-principles accuracy, remains unresolved. Xu et al.~\cite{r42}
investigated ITC in lateral MoS\textsubscript{2}/\allowbreak{}MoSe\textsubscript{2} heterostructures using an
empirical Stillinger--Weber potential, but their study focused on
interface defect type rather than crystallographic orientation.
Moreover, their simulations became numerically unstable at 500 K. Liu et
al.~\cite{r39} developed a DeePMD potential for the related MoS\textsubscript{2}/\allowbreak{}WS\textsubscript{2}
system using a training dataset that included both armchair- and
zigzag-oriented interfaces; however, their ITC calculations considered
only a single orientation, leaving the orientation dependence of ITC and
its underlying phonon mechanism unexplored.

To address this gap, we developed a dedicated neural network potential
for monolayer MoS\textsubscript{2}, MoSe\textsubscript{2}, and their lateral heterostructures using the
deep potential molecular dynamics (DeePMD) framework~\cite{r34}. The
resulting MLIP was benchmarked against first-principles calculations,
including ab initio molecular dynamics (AIMD), to verify its ability to
reproduce the relevant structural and vibrational properties without
overfitting to the training data. The validated MLIP was then employed
in molecular dynamics simulations to calculate the ITC of lateral
MoS\textsubscript{2}/\allowbreak{}MoSe\textsubscript{2} heterostructures for both zigzag and armchair interface
orientations. By comparing these two configurations and analyzing their
vibrational characteristics, we establish crystallographic orientation
as a quantitative parameter governing ITC and elucidate its underlying
phonon mechanism.

The remainder of this paper is organized as follows. Section 2 describes
the construction of the training dataset and the DLP training procedure.
Section 3 validates the trained potential against first-principles
benchmarks and presents the resulting ITC and phonon-based analysis for
the two interface orientations. Section 4 summarizes the key findings
and their implications for thermal management in 2D lateral
heterostructure devices.

\FloatBarrier
\section{Computational details}

\subsection{Dataset construction}

\begin{figure*}[!tbp]
  \centering
  \renewcommand{\imgH}{0.90\textheight}
  \img[1.00\linewidth]{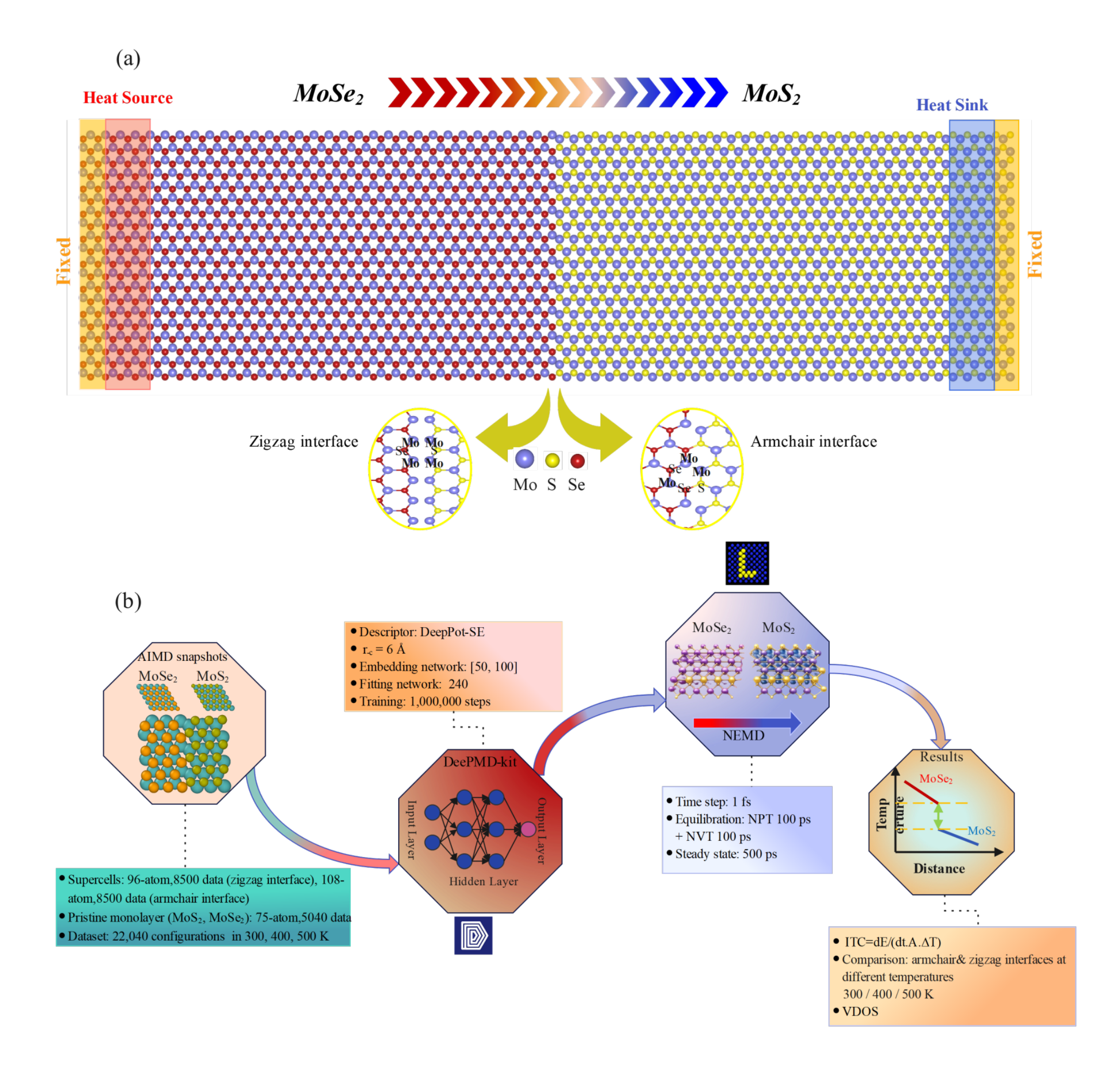}
  \caption{(a) Schematic of the NEMD simulation setup for calculating interfacial thermal conductance in the MoSe\textsubscript{2}/\allowbreak{}MoS\textsubscript{2} heterostructure (zigzag and armchair interfaces), and (b) the workflow for constructing the machine-learning potential (DeePMD-kit) used to compute ITC.}
  \label{fig:1}
\end{figure*}

Since the performance of an MLIP is fundamentally constrained by the
quality of its training data, considerable effort was devoted to
constructing a dataset that comprehensively captures the range of
structures and interfaces relevant to this system. The starting point
was AIMD simulations of monolayer MoS\textsubscript{2} and MoSe\textsubscript{2}, plus their lateral
heterostructures --- these snapshots constituted the structural
foundation of the dataset. Both monolayers crystallize in a hexagonal
lattice, space group P-6m2 (187), and the lateral MoSe\textsubscript{2}/\allowbreak{}MoS\textsubscript{2}
heterostructure was built by joining the two monolayers edge-to-edge
within the same atomic plane. We looked at two interface orientations
here --- armchair and zigzag --- since these are the two high-symmetry
directions along which the MoS\textsubscript{2} and MoSe\textsubscript{2} domains can naturally meet
(Fig.~\ref{fig:1}a). The optimized lattice parameters of the primitive unit cells
were determined to be \(a = 3.192\)Å for monolayer MoS\textsubscript{2}, and
\(a = 3.322\)Å for monolayer MoSe\textsubscript{2}. These values are in good agreement
with previously reported theoretical and experimental results,
confirming the reliability of the optimized structures~\cite{r43,r44,r45}.

The heterostructure part of the dataset required particular attention,
as interfacial thermal transport is highly sensitive to interface
geometry. To avoid training an MLIP that only ``understands'' one
interface type, we built sharp-interface MoS\textsubscript{2}/\allowbreak{}MoSe\textsubscript{2} heterostructures
along both the zigzag and armchair directions --- a 96-atom supercell
for the zigzag case and a 108-atom supercell for the armchair one. These
two supercells became the starting configurations for the AIMD runs that
fed the training set.

To construct commensurate interfaces, the in-plane dimensions of the
MoS\textsubscript{2} and MoSe\textsubscript{2} domains were carefully selected to minimize the lattice
mismatch while preserving the periodicity and structural compatibility
of the interface. The residual mismatch was accommodated by applying an
equal and symmetric strain to both domains in the armchair and zigzag
configurations.

On top of that, we added 12 more structures generated through the DP-GEN
active-learning framework~\cite{r46,r47,r48} to further expand the
configurational space and ensure adequate representation of less common
atomic arrangements. All DFT calculations relied on the generalized
gradient approximation with the PBE exchange-correlation functional~\cite{r49}. Owing to the high computational cost of these
calculations, Brillouin-zone sampling was restricted to the $\Gamma$-point
only, and a 336-eV cutoff was used for the plane-wave basis.

SCF calculations were treated as converged once the total energy between
iterations changed by less than 1×10\textsuperscript{$-$4} eV per atom. In total, the
dataset used to train the neural network potential contains 22,040
configurations, spread across three temperatures --- 300, 400, and 500
K. It breaks down into three chunks: 5,040 configurations from pristine
monolayer MoS\textsubscript{2} and MoSe\textsubscript{2}, 8,500 from the armchair-interface
heterostructure, and another 8,500 from the zigzag one. We set it up
this way on purpose, so both interface geometries would be well
represented across the temperature range, enabling the potential to
capture the structural and vibrational behavior that drives interfacial
thermal transport in these systems. Before training, we split the
dataset randomly --- 90\% for training, the remaining 10\% held back as
a test set to check predictive accuracy. This split is essential for
verifying that the potential generalizes rather than memorizing the
training data, and it supports the reliability of its predictions for
both armchair and zigzag interfaces across the whole temperature range.

\subsection{Training procedure}

We trained the deep learning potential using the DeePMD-kit framework.
Deep neural networks tend to outperform more traditional machine
learning methods here, largely because they\textquotesingle re good at
picking up on subtle, symmetry-preserving patterns buried in large AIMD
datasets~\cite{r50}. The whole framework rests on one core assumption: the
total energy of a system of N atoms can be broken down into
contributions from each atom\textquotesingle s local environment:

\begin{equation}
  \adjustbox{max width=\columnwidth}{$\displaystyle E = \sum_{i}^{}E_{i}$}
\end{equation}

\noindent where $E$\textsubscript{i} denotes the atomic energy contribution of atom i, which is
determined entirely by the local atomic environment surrounding it.

A central challenge in constructing an MLIP lies in representing the
local atomic environment in a way that still respects translational,
rotational, and permutational symmetry~\cite{r51}. To handle this, we used
the Smooth Edition of Deep Potential (DeepPot-SE)~\cite{r52}, which
converts raw atomic coordinates into a symmetry-preserving descriptor
space that the neural network can actually train on. To keep the
resulting potential energy surface smooth and differentiable, we set the
cutoff radius rc at 6 Å and the smoothing cutoff rcs at 5.50 Å --- this
way atomic interactions fade out gradually near the boundary instead of
just cutting off abruptly. The hyperbolic tangent function was used as
the activation function throughout.

For the network architecture, the embedding network used layer sizes of[50, 100], and the fitting network had three hidden layers of 240
neurons each. Training ran for 1,000,000 steps, with the learning rate
decaying exponentially from 0.001 down to 3.51 × 10\textsuperscript{$-$8} --- so the model
could take big steps early on and then fine-tune as it approached
convergence. The loss function L combined both energy and force errors
during training~\cite{r34}:

\begin{equation}
  \adjustbox{max width=\columnwidth}{$\displaystyle L\left( p_{e},p_{f} \right) = p_{e}{\mathrm{\Delta}e}^{2} + \frac{p_{f}}{3N}\sum_{i}^{}{|{\mathrm{\Delta}F}_{i}|}^{2}$}
\end{equation}

In this expression, $\Delta$ denotes the deviation between the reference
training data and the predictions generated by the DLP model, while
\(N\) corresponds to the total number of atoms in the simulated system.
The quantity \(e\) represents the energy per atom, and \(F_{i}\) denotes
the force acting on the \(i\_ th\) atom. The coefficients \(p_{e}\) and
\(p_{f}\ \)govern the relative weighting of the energy and force
contributions within the loss function. These parameters were carefully
tuned to strike an appropriate balance between accurately reproducing
the total energy and correctly capturing atomic forces, thereby ensuring
that the resulting neural network potential faithfully reflects both the
equilibrium (static) and time-dependent (dynamic) behavior of the
system.

Fig.~\ref{fig:1}b lays out the overall workflow: build the AIMD dataset, train the
MLIP with DeePMD-kit, run NEMD simulations in LAMMPS, and finally
extract the ITC of the heterostructure.

\FloatBarrier
\section{Results and discussion}

\subsection{Validation of the Trained DeePMD Potential}

\begin{figure*}[!tbp]
  \centering
  \renewcommand{\imgH}{0.90\textheight}
  \img[1.00\linewidth]{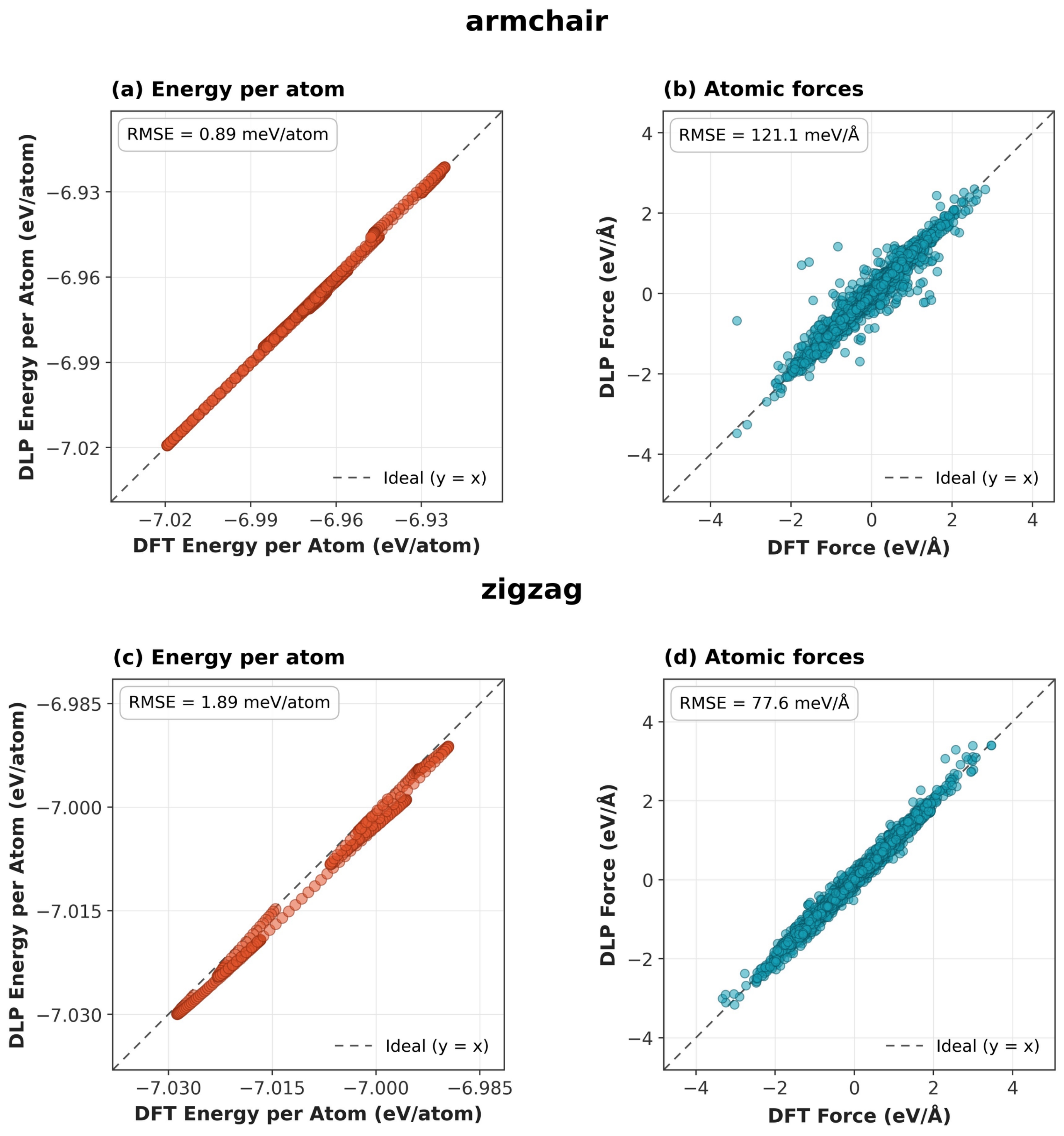}
  \caption{Correlation between DLP-predicted and DFT-calculated energies and atomic forces for the held-out test set of the MoSe\textsubscript{2}/\allowbreak{}MoS\textsubscript{2} interface: (a) energy and (b) force correlation for the armchair interface; (c) energy and (d) force correlation for the zigzag interface.}
  \label{fig:2}
\end{figure*}

To assess the accuracy of the trained model, we compared DLP-predicted
energies and forces with DFT reference values for the 10\% test set
excluded from training. Fig.~\ref{fig:2}(a, b) shows this comparison for the
armchair interface, and Fig.~\ref{fig:2}(c, d) shows the corresponding comparison
for the zigzag interface. Both orientations show close agreement along
the parity line. The energy and force RMSEs are 0.89 meV/\allowbreak{}atom and 121.1
meV/Å for the armchair interface, and 1.89 meV/\allowbreak{}atom and 77.6 meV/Å for
the zigzag interface.

\begin{figure*}[!tbp]
  \centering
  \renewcommand{\imgH}{0.90\textheight}
  \img[1.00\linewidth]{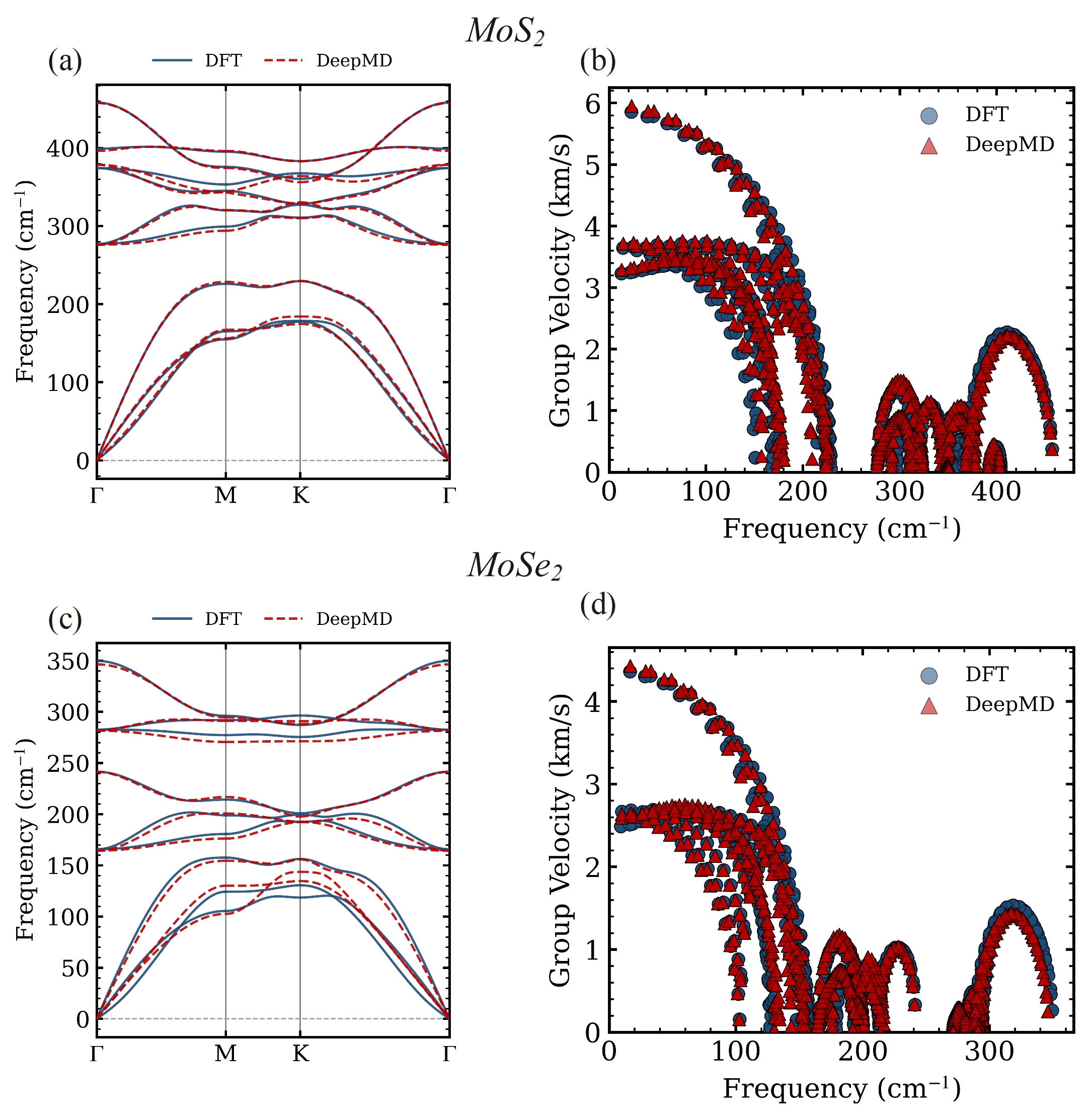}
  \caption{Phonon dispersion curves and group velocities for MoS\textsubscript{2} (a, b) and MoSe\textsubscript{2} (c, d), comparing DFT and DeePMD results.}
  \label{fig:3}
\end{figure*}

To further assess the reliability of the trained MLIP beyond energy and
force predictions, we also computed phonon dispersion relations for
pristine MoS\textsubscript{2} and MoSe\textsubscript{2} monolayers and compared them with DFT, as shown
in Figs.~\ref{fig:3}(a) and 3(c). The DeePMD phonon branches follow the DFT curves
closely across the Brillouin zone for both monolayers. The agreement is
nearly perfect for MoS\textsubscript{2}, whereas for MoSe\textsubscript{2} small deviations are visible
in some low-frequency and optical branches, mainly between the M and K
points. Despite these small deviations, the phonon group velocities
(Figs.~\ref{fig:3}(b) and 3(d)) remain in close agreement with DFT across the
entire frequency range for both monolayers. Since phonon group velocity
is a key quantity governing lattice heat transport, this level of
agreement supports the ability of the trained potential to capture the
vibrational and phonon transport physics of both materials, which in
turn gives us confidence in using it for the NEMD simulations of
interfacial thermal transport in the MoSe\textsubscript{2}/\allowbreak{}MoS\textsubscript{2} heterostructure that
follow. These results are also in good agreement with previously
published phonon dispersion data for monolayer MoS\textsubscript{2} and MoSe\textsubscript{2}~\cite{r16,r24}.

To evaluate the ITC of the MoSe\textsubscript{2}/\allowbreak{}MoS\textsubscript{2} lateral heterostructure, NEMD
simulations were employed~\cite{r53,r54}.

\subsection{\texorpdfstring{Interfacial Thermal Conductance of the MoSe\textsubscript{2}/\allowbreak{}MoS\textsubscript{2} Heterostructure}{Interfacial Thermal Conductance of the MoSe2/\allowbreak{}MoS2 Heterostructure}}

All simulations were carried out using the Large-scale Atomic/\allowbreak{}Molecular
Massively Parallel Simulator (LAMMPS)~\cite{r55}, with the trained
DLP handling the atomic interactions throughout.

Both zigzag and armchair interface configurations were examined this way
(Fig.~\ref{fig:1}a). The armchair heterostructure had a simulation cell
measuring 20.93 nm × 5.64 nm along x and y, while the zigzag one came in
at 21.45 nm × 5.86 nm. In both cases, we left a 3 nm vacuum gap along
the out-of-plane (z) direction so the periodic images of the monolayer
wouldn\textquotesingle t interact with each other. We ran simulations at
300 K, 400 K, and 500 K, with a 1 fs integration time step.

Once the simulation cell and initial atomic configuration were set up,
we ran a short energy minimization first to eliminate unphysical atomic
overlaps or unfavorable local configurations before starting the actual
dynamics. From there, the system was equilibrated in the NPT ensemble at
300 K for 100 ps, letting both temperature and pressure settle into
their target states. After that, we ran a second equilibration stage ---
this time in the NVT ensemble, also at 300 K for 100 ps --- to
equilibrate the velocity distribution and ensure thermal stability
before moving on to the production run.

This method was applied identically at all three target mean
temperatures (300, 400, and 500 K), with NPT and NVT equilibration
carried out at each respective temperature for 100 ps. During NPT
equilibration, pressure was controlled independently along the x, y, and
z directions (aniso coupling) at a target of 1 bar with a barostat
damping constant of 5 ps; a thermostat damping constant of 0.1 ps was
used in both the NPT and NVT stages.

For the non-equilibrium heat transport part, we split the simulation
cell along the transport (X) direction into four regions: fixed, hot,
cold, and active. At each end of the cell, we set a 5 Å-thick fixed
region where atoms were held in place to prevent unphysical drift of the
system.

Adjacent to these fixed boundaries, the hot and cold thermostat regions
were placed, held at T\textsubscript{0}+20 K and T\textsubscript{0}$-$20 K, respectively, using a
Nosé--Hoover thermostat with a damping constant of 0.1 ps --- this fixed
40 K temperature difference is what drives a steady heat flux from the
hot side to the cold side, applied consistently across all three mean
temperatures investigated: 320/280 K at 300 K, 420/380 K at 400 K, and
520/480 K at 500 K.

The active region in between, where we actually track heat flow, was
split further into 20 equally spaced slabs so we could resolve the
temperature profile spatially. Each NEMD run went for 500 ps to let the
system settle into a non-equilibrium steady state, then another 1000 ps
of production run to collect the thermal and structural data.

Four independent NEMD simulations, each initialized with a different
random velocity seed, were performed for a representative case (the
armchair interface at 300 K) to quantify statistical uncertainty; the
resulting relative uncertainty was applied uniformly across the other
interface orientations and temperatures.

\begin{figure*}[!tbp]
  \centering
  \renewcommand{\imgH}{0.90\textheight}
  \img[1.00\linewidth]{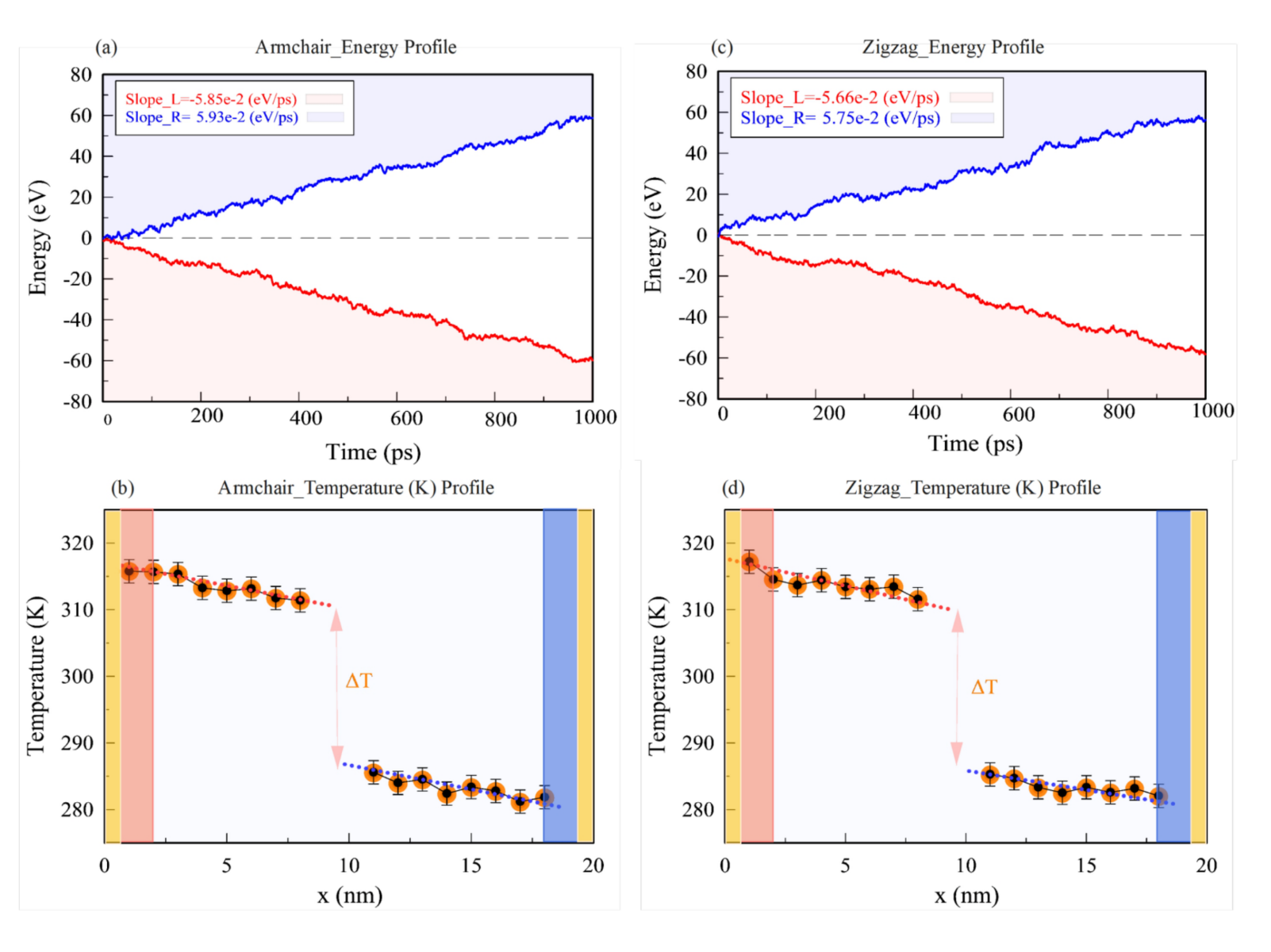}
  \caption{(a) and (c) display the cumulative energy profiles of the thermostats during the NEMD simulations for the armchair and zigzag MoSe\textsubscript{2}/\allowbreak{}MoS\textsubscript{2} interfaces, respectively; Fig.~\ref{fig:4}(b) and (d) show the corresponding steady-state temperature profiles for the armchair and zigzag interfaces.}
  \label{fig:4}
\end{figure*}

Fig.~\ref{fig:4}(a) and (c) show how energy builds up at the hot and cold
thermostats for the armchair and zigzag interfaces over that 1000 ps
production run. The fact that this accumulation is linear and symmetric
at both the source and the sink tells us a steady-state heat flux had
been reached --- the slopes of these curves were used to determine the
heat transfer rate for each interface.

Fig.~\ref{fig:4}(b) and (d) show the corresponding temperature profiles. Each
one has two linear regions, one within the MoSe\textsubscript{2} domain and one within
MoS\textsubscript{2}, reflecting each material\textquotesingle s own thermal
conductivity, plus a temperature jump ($\Delta$T) right at the interface ---
this comes from phonon scattering caused by the mismatch in lattice
structure and vibrational properties between the two materials. From
heat transfer rate and this temperature discontinuity, we calculated the
ITC. The fact that both the heat flux and the temperature jump differ
between the armchair and zigzag interfaces tells us that interface
orientation actually matters for phonon transport, and by extension, for
how heat crosses the MoSe\textsubscript{2}/\allowbreak{}MoS\textsubscript{2} interface.

The ITC was then calculated using the following relation~\cite{r36,r56,r57}:

\begin{equation}
  \adjustbox{max width=\columnwidth}{$\displaystyle ITC = \frac{dE}{dt.A.\mathrm{\Delta}T}$}
\end{equation}

\noindent where $\Delta$T is the temperature discontinuity across the interface,
\(\frac{dE}{dt}\) is the rate at which the thermostatted regions
exchange energy, and \(A\) is the cross-sectional area of the interface
through which heat is transported. The cross-sectional area A was
calculated as the product of the in-plane cell width and an effective
monolayer thickness of 3.32 Å, measured as the vertical span between the
outermost chalcogen atoms across the Mo plane in the relaxed monolayer
structure.

\begin{figure*}[!tbp]
  \centering
  \renewcommand{\imgH}{0.90\textheight}
  \img[0.60\linewidth]{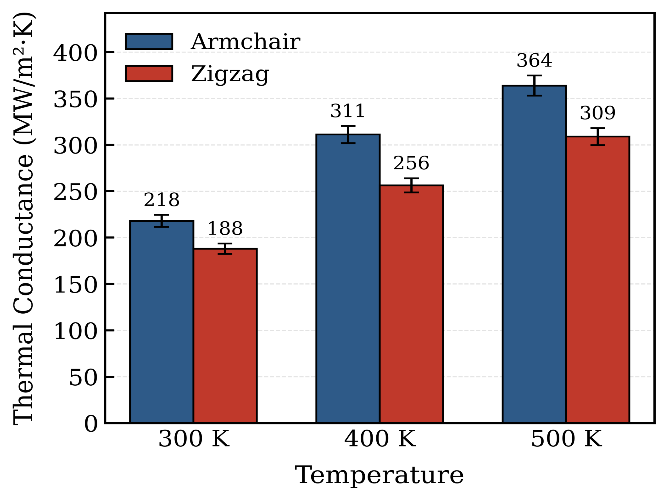}
  \caption{ITC of the armchair and zigzag MoSe\textsubscript{2}/\allowbreak{}MoS\textsubscript{2} interfaces at 300, 400, and 500 K.}
  \label{fig:5}
\end{figure*}

We looked at ITC values for both interface orientations at 300, 400, and
500 K, shown in Fig.~\ref{fig:5}. Across all three temperatures, the armchair
interface comes out consistently higher than the zigzag one --- 218 vs.
188 MW/\allowbreak{}m\textsuperscript{2}·K at 300 K, 311 vs. 256 MW/\allowbreak{}m\textsuperscript{2}·K at 400 K, and 364 vs. 309
MW/\allowbreak{}m\textsuperscript{2}·K at 500 K. This orientation dependence suggests phonons cross the
interface more efficiently along the armchair direction. Structural
analysis shows that the two domains are bridged by direct S--Se bonds
across the seam. The mean S--Se bond length is 2.25 Å for the armchair
interface, close to the equilibrium S--Se bond length
(\textasciitilde2.2--2.25 Å), compared to 2.47 Å (approximately 10\%
longer) for the zigzag interface, reflecting an intrinsic geometric
consequence of the interface orientation.

For both orientations, ITC climbs steadily with temperature --- by about
67\% for armchair and 64\% for zigzag between 300 and 500 K.

This is consistent with stronger anharmonic phonon-phonon scattering at
higher temperatures: larger vibrational amplitudes push the system
further into the anharmonic part of the interatomic potential,
strengthening phonon-phonon coupling and opening additional pathways for
energy transfer across the interface. Similar increases in ITC with
temperature have been reported for other TMD and solid--solid interfaces~\cite{r42,r58,r59}.

The two orientations show comparable temperature sensitivity, with
armchair exhibiting a marginally larger relative increase. In absolute
terms, the ITC gap between armchair and zigzag widens from
\textasciitilde30 MW/\allowbreak{}m\textsuperscript{2}·K at 300 K to \textasciitilde55 MW/\allowbreak{}m\textsuperscript{2}·K at 400
K, and remains essentially unchanged up to 500 K.

\subsection{Finite-size effects}

The limited size of the simulation cell places an intrinsic constraint
on the maximum phonon mean free path (MFP) that can be captured,
making it essential to account for finite-size effects.

\begin{figure*}[!tbp]
  \centering
  \renewcommand{\imgH}{0.50\textheight}
  \img[1.00\linewidth]{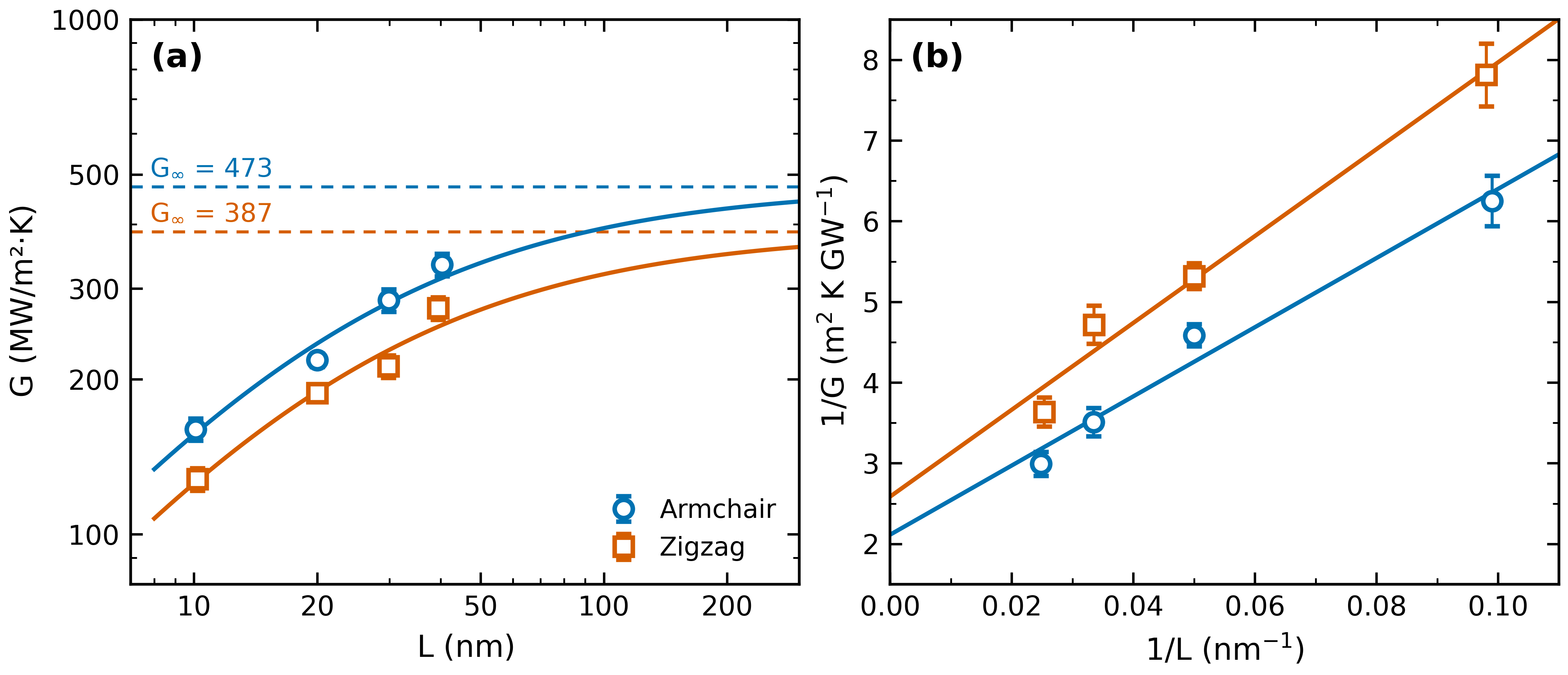}
  \caption{(a) Dependence of the ITC of the armchair and zigzag MoSe\textsubscript{2}/\allowbreak{}MoS\textsubscript{2} interfaces on the length at 300 K; (b) Inverse ITC (1/G) as a function of the inverse length (1/L).}
  \label{fig:6}
\end{figure*}

Figure~\ref{fig:6}a illustrates the dependence of the ITC of the armchair and
zigzag MoSe\textsubscript{2}/\allowbreak{}MoS\textsubscript{2} interfaces on the cell length at 300 K. With the width
fixed at 5.64 nm (armchair) and 5.86 nm (zigzag), the ITC was calculated
for total cell lengths of 10, 20, 30, and 40 nm. As the length
increases, the ITC of the armchair interface rises from about 160 to 334
MW m\textsuperscript{$-$2} K\textsuperscript{$-$1} and that of the zigzag interface from about 128 to 275 MW m\textsuperscript{$-$2}
K\textsuperscript{$-$1}, roughly a factor of two, and neither curve has saturated at 40 nm.
When the domain length is smaller than the phonon MFP, the ITC increases
with length, and the relationship between them is described as follows~\cite{r60,r61}:

\begin{equation}
  \adjustbox{max width=\columnwidth}{$\displaystyle \frac{1}{G (L) } = \frac{1}{G (\infty) } + \frac{C}{ L}$}
\end{equation}

\noindent where G(L) denotes the ITC of a cell with length L, G ($\infty$) represents the
ITC at infinite length, and C is a fitting constant. As shown in Fig.~\ref{fig:6}b, 1/G is close to linear in 1/L for both orientations, and
R\textsuperscript{2} =0.98. The fits give G ($\infty$) = 473 MW m\textsuperscript{$-$2} K\textsuperscript{$-$1}
for armchair and 387 MW m\textsuperscript{$-$2} K\textsuperscript{$-$1} for zigzag. The armchair interface has a
higher ITC than the zigzag interface at every length.

The transverse period was kept fixed, so our results refer to coherent
interfaces with these periods, which are in the range of widths used in
NEMD studies of lateral heterostructures with machine-learning
potentials~\cite{r39,r62}.

\subsection{VDOS and overlap factor (S) analysis}

To further elucidate the role of interface geometry, we computed the
VDOS for atoms within 0.5 nm of the interface --- the region that
matters most for interfacial thermal transport --- for both the armchair
and zigzag setups. The VDOS was obtained from the Fourier transform of
the velocity autocorrelation function, and comparing the spectra between
the two orientations shows how the interface geometry shapes the local
vibrational spectrum, and in turn the phonon overlap that governs how
heat gets across the interface~\cite{r63}.

\begin{equation}
  \adjustbox{max width=\columnwidth}{$\displaystyle VDOS(\omega) = \frac{1}{\sqrt{2\pi}}\int_{- \infty}^{\infty}e^{- i\omega t}\frac{\left\langle v(t).v(0) \right\rangle}{\left\langle v(0).v(0) \right\rangle}dt$}
\end{equation}

\begin{figure*}[!tbp]
  \centering
  \renewcommand{\imgH}{0.50\textheight}
  \img[1.00\linewidth]{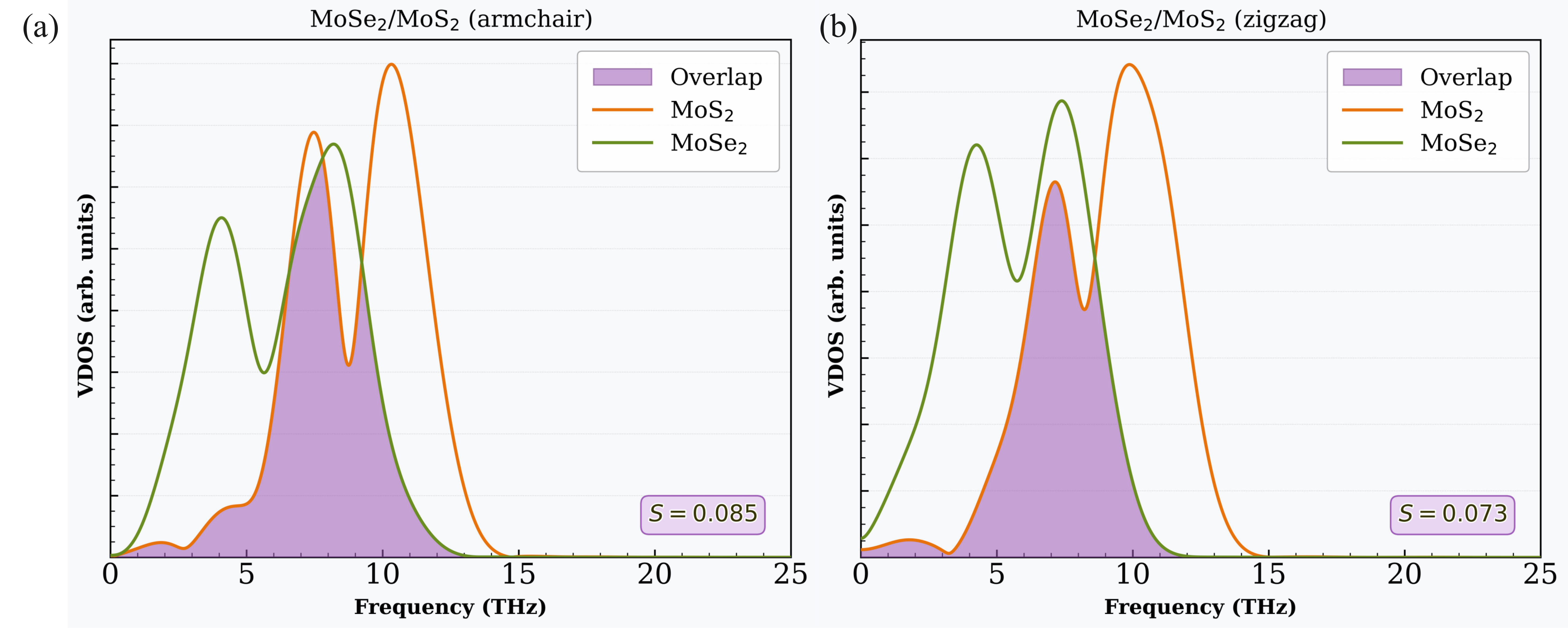}
  \caption{VDOS spectra near the interface for (a) the armchair MoSe\textsubscript{2}/\allowbreak{}MoS\textsubscript{2} interface and (b) the zigzag MoSe\textsubscript{2}/\allowbreak{}MoS\textsubscript{2} interface at 300 K.}
  \label{fig:7}
\end{figure*}

\noindent where VDOS($\omega$) denotes the vibrational density of states at frequency $\omega$,
v(t) and v (0) represent the atomic velocity vectors at time t and at
the reference time 0, respectively, and the angle brackets $\langle$\ldots$\rangle$
denote an ensemble average taken over all atoms within the selected
interfacial region and over multiple time origins.The VDOS spectra shown
in Fig.~\ref{fig:7} were computed at 300 K.

To quantify the vibrational compatibility across the interface, we
calculated a spectral overlap factor (S) between the VDOS spectra of
MoSe\textsubscript{2} and MoS\textsubscript{2}, following the formulation of Ref.~\cite{r59}.

\begin{equation}
  \adjustbox{max width=\columnwidth}{$\displaystyle S = \frac{\int_{}^{}{{VDOS}_{{MoSe}_{2}}(\omega){VDOS}_{{MoS}_{2}}(\omega)}d\omega}{\int_{}^{}{{VDOS}_{{MoSe}_{2}}(\omega)d\omega}\int_{}^{}{{VDOS}_{{MoS}_{2}}(\omega)}d\omega}$}
\end{equation}

\noindent where \(VDOS_{MoSe_{2}}(\omega)\)and \(VDOS_{MoS_{2}}(\omega)\)represent
the vibrational density of states of MoSe\textsubscript{2} and MoS\textsubscript{2}, respectively.

Figures~\ref{fig:7}(a) and 7(b) show the VDOS spectra for the armchair
and zigzag MoSe\textsubscript{2}/\allowbreak{}MoS\textsubscript{2} interfaces. The MoSe\textsubscript{2} spectrum extends to lower
frequencies in both cases, with peaks near 4 and 8 THz. In contrast, the
MoS\textsubscript{2} spectrum shifts to higher frequencies, with peaks in the 7--11 THz
range. This shift is attributable to the vibrational mismatch between
the two materials, arising from the difference in mass and bond
stiffness between S and Se atoms.

The spectral overlap factor S, computed via Eq. (6), provides a
quantitative measure of the vibrational compatibility between the two
materials, where larger values correspond to stronger vibrational
overlap across the interface. The shaded purple region in Fig.~\ref{fig:7} serves
as a qualitative visual guide to the frequency range where the two
spectra co-occur.

For the armchair interface (Fig.~\ref{fig:7}(a)), S is 0.085, while for the zigzag
interface (Fig.~\ref{fig:7}(b)) it is lower, 0.073, which is qualitatively
consistent with the higher ITC of the armchair interface at 300 K. Since
S only measures same-frequency (elastic) vibrational compatibility and
was evaluated at 300 K only, we regard it as a qualitative indicator
rather than a quantitative predictor of ITC. The persistence of the
armchair advantage at 400 and 500 K, where inelastic (anharmonic)
channels may become more important, suggests that contributions beyond
elastic spectral matching, such as the interfacial bonding differences
discussed in Sec. 3.2, may also play a role.

\begin{table*}[!tbp]
  \centering
  \caption{Comparison of room-temperature ITC values collected from prior studies on several typical 2D heterostructures.}
  \label{tab:1}
  \footnotesize
  \begin{adjustbox}{max width=\linewidth}
\begin{tabular}{@{}lllll@{}}
\toprule
Materials & Ref. & Method & Interface geometry & ITC (MW/\allowbreak{}m\textsuperscript{2}. k) \\
\midrule
MoS\textsubscript{2}/\allowbreak{}WS\textsubscript{2} &~\cite{r39} X. Liu et al. & MLIP & Lateral & 464.89 \\
MoS\textsubscript{2}/\allowbreak{}\emph{h}-BN &~\cite{r64} Y. Liu et al. & Raman spectroscopy & Vertical (vdW-stacked) & 17.0 ± 0.4 \\
Graphene/\allowbreak{}MoS\textsubscript{2} &~\cite{r65} S. Vaziri et al. & Raman spectroscopy & Vertical (vdW-stacked) & 12.048 \\
BP/\allowbreak{}MoS\textsubscript{2} &~\cite{r66} B. Wu et al. & MD & Vertical (vdW-stacked) & 5.572 \\
MoSe\textsubscript{2}/\allowbreak{}MoS\textsubscript{2}(Armchair) & This work & MLIP & Lateral & 473 \\
MoSe\textsubscript{2}/\allowbreak{}MoS\textsubscript{2}(Zigzag) & This work & MLIP & Lateral & 387 \\
\bottomrule
\end{tabular}
  \end{adjustbox}
\end{table*}

Table~\ref{tab:1} compares the room-temperature ITC values obtained in this study
with those reported for other 2D heterostructures. Our MoSe\textsubscript{2}/\allowbreak{}MoS\textsubscript{2}
numbers --- 473 MW/\allowbreak{}m\textsuperscript{2}·K for armchair, 387 for zigzag (extrapolated to
infinite cell length) --- are substantially higher than previously
reported values for systems such as MoS\textsubscript{2}/\allowbreak{}h-BN (17.0 ± 0.4 MW/\allowbreak{}m\textsuperscript{2}·K),
graphene/\allowbreak{}MoS\textsubscript{2} (12.048 MW/\allowbreak{}m\textsuperscript{2}·K), or BP/\allowbreak{}MoS\textsubscript{2} (5.572 MW/\allowbreak{}m\textsuperscript{2}·K). Those
systems all rely on weak interlayer coupling, so phonon transmission is
limited from the start. This contrast underscores the fundamentally
different transport mechanism at a covalently bonded lateral interface,
where direct atomic bonds across the junction give phonons a much more
efficient path than the van der Waals coupling in vertically stacked
structures.

We note that MoS\textsubscript{2}/\allowbreak{}h-BN, graphene/\allowbreak{}MoS\textsubscript{2}, and BP/\allowbreak{}MoS\textsubscript{2} are vertically
stacked (van der Waals) heterostructures, for which the interfacial
cross-sectional area corresponds to the full, in-plane sample area. In
contrast, MoS\textsubscript{2}/\allowbreak{}WS\textsubscript{2}~\cite{r39} and the present MoSe\textsubscript{2}/\allowbreak{}MoS\textsubscript{2} system are
lateral heterostructures, for which the cross-sectional area depends on
an assumed effective monolayer thickness (3.32 Å in this work).

On the other hand, the lateral MoS\textsubscript{2}/\allowbreak{}WS\textsubscript{2} interface reported using an
MLIP-based approach reaches 464.89 MW/\allowbreak{}m\textsuperscript{2}·K for a cell length of about 20
nm. At a comparable length (L $\approx$ 21 nm), our MoSe\textsubscript{2}/\allowbreak{}MoS\textsubscript{2} interface gives
218 MW/\allowbreak{}m\textsuperscript{2}·K (armchair) and 188 MW/\allowbreak{}m\textsuperscript{2}·K (zigzag), roughly half of this
value, whereas the extrapolated infinite-length values listed in Table~\ref{tab:1}
are of the same order. This discrepancy is likely attributable to
differences in lattice mismatch, bonding strength, and how well the
vibrational spectra of the two monolayers overlap. Taken together, these
comparisons suggest that, within lateral TMD/\allowbreak{}TMD heterostructures, the
choice of constituent materials strongly influences the resulting ITC.

\FloatBarrier
\section{Conclusions}

In this work, we built a deep learning interatomic potential using
DeePMD-kit and paired it with non-equilibrium MD simulations to look at
ITC in lateral MoSe\textsubscript{2}/\allowbreak{}MoS\textsubscript{2} heterostructures, comparing armchair and
zigzag interface orientations. The trained potential matches DFT
reference energies, forces, and phonon properties for pristine MoS\textsubscript{2} and
MoSe\textsubscript{2} quite well, which gives us confidence in using it for large-scale
simulations of the heterostructure. The dependence of the ITC on cell
length was examined for lengths from 10 to 40 nm, and extrapolation to
infinite length gives 473 MW/\allowbreak{}m\textsuperscript{2}·K for the armchair and 387 MW/\allowbreak{}m\textsuperscript{2}·K for
the zigzag interface. The armchair interface shows a higher ITC than the
zigzag one at every length considered (at 300 K) and at all temperatures
studied (at L $\approx$ 21 nm). For both orientations, ITC increases with
temperature, which is consistent with enhanced anharmonic phonon--phonon
scattering. Analysis of the vibrational density of states at 300 K shows
more overlap of vibrational modes of similar frequencies at the armchair
interface (S = 0.085) compared to the zigzag interface (S = 0.073). All
in all, the findings suggest that the direction of the interface is an
important factor for ITC, consistent with the greater vibrational
compatibility between the two lattices in the armchair configuration.
Collectively, this study establishes a machine-learning-based framework
for studying thermal transport in lateral TMD heterostructures, and
shows that interface orientation is a usable lever for tuning ITC in
coherent MoSe\textsubscript{2}/\allowbreak{}MoS\textsubscript{2} junctions.

\FloatBarrier

\section*{Conflicts of interest}

There are no conflicts to declare.

\FloatBarrier

\section*{Acknowledgements}

This work is based upon research funded by Iran National Science
Foundation (INSF) under project No.4040188.

\FloatBarrier

\section*{Data availability statement}

The data that support the findings of this study are openly available at
the following URL/\allowbreak{}DOI:

www.aissquare.com/\allowbreak{}datasets/\allowbreak{}detail?pageType=datasets\&id=432.

\FloatBarrier

\end{document}